\documentclass[
  journal=largetwo,
  manuscript=article,
]{cup-journal}

\usepackage{amsmath}
\usepackage[nopatch]{microtype}
\usepackage{booktabs}
\usepackage{microtype}

\newcommand{\kms}{km\,s$^{-1}$}

\newcommand{\df}[1]{NGC\ 1052-DF{#1}}
\newcommand{\vsys}{v_{\rm sys}}

\title{Rotation of the Globular Cluster Population of the Dark Matter Deficient Galaxy \df4: 
Implication for the total mass}

\author{Yuan (Cher) Li}
\affiliation{Department of Statistics, The University of Auckland, Private Bag 92019, Auckland 1142, New Zealand}
\email[Y. Li]{yli464@aucklanduni.ac.nz}

\author{Brendon J. Brewer}
\affiliation{Department of Statistics, The University of Auckland, Private Bag 92019, Auckland 1142, New Zealand}

\author{Geraint F. Lewis}
\affiliation{Sydney Institute for Astronomy, School of Physics, A28, The University of Sydney, NSW 2006, Australia}

\begin{document}

\begin{abstract}
We explore the globular cluster population of \df4, a dark matter deficient galaxy, using Bayesian inference to search for the presence of rotation. The existence of such a rotating component is relevant to the estimation of the mass of the galaxy, and therefore the question of whether \df4 is truly deficient of dark matter,similar to \df2, another galaxy in the same group. The rotational characteristics of seven globular clusters in \df4 were investigated, finding that a non-rotating kinematic model has a higher Bayesian evidence than a rotating model, by a factor of approximately 2.5. In addition, we find that under the assumption of rotation, its amplitude must be small. This distinct lack of rotation strengthens the case that, based on its intrinsic velocity dispersion, \df4 is a truly dark matter deficient galaxy.
\end{abstract}

\section{Introduction}
\label{introduction}


Ultra-diffuse galaxies (UDGs) are defined by their extremely low matter densities and luminosities, with a surface brightness from 25 to 28 mag arcsec$^{-2}$ and a relatively large scale radius of about $ > 1.5$ kpc \citep{2015ApJ...798L..45V,2015ApJ...807L...2K}. According to \citet{2016ApJ...828L...6V}, some UDGs appear to have very high dark matter fractions with very little stellar material.
Alternatively, other UDGs appear to possess much less dark matter than expected \citep[e.g.][]{2018Natur.555..629V}. 

NGC 1052-DF2 is one such claimed matter deficient UDG in the NGC 1052 group. \citet{2018ApJ...864L..18V} first reported that NGC1052-DF2 has a comparatively small velocity dispersion of about 3.2 \kms, which is the basis for the galaxy's low mass estimate, and hence implied deficiency of dark matter. This finding prompted controversy since dark matter is thought to be essential to galaxy formation and evolution and presumed to be significant in almost all galaxies. 
Through a Bayesian-based analysis,
\citet{2018ApJ...859L...5M} concluded that NGC 1052-DF2’s inherent velocity dispersion is about 9.5 \kms, which somewhat mitigates the shortage of dark matter. However, \citet{2019ApJ...874L...5V} discovered a second galaxy in the NGC 1052 group, naming it NGC 1052-DF4. This galaxy is also a UDG and is very similar to the NGC 1052-DF2 galaxy in size, surface brightness, and shape. \citet{2019ApJ...874L...5V} found that NGC 1052-DF4 also possessed a relatively low velocity dispersion, and hence concluded it is also dark matter deficient.

The potential lack of dark matter in NGC 1052-DF2 and NGC 1052-DF4 has prompted investigation from several angles. Some have focused on determining the distance to these galaxies \citep{2019MNRAS.486.1192T,2020ApJ...895L...4D,2021ApJ...909..179S,2021MNRAS.504.1668Z}, with the possibility that a smaller distance would bring the properties of these galaxies inline with the overall galaxy populations. Others have considered the estimation of their intrinsic velocity dispersion \citep{2018ApJ...864L..18V,2018ApJ...859L...5M,2019A&A...625A..77F,2019ApJ...874L...5V}, as well as the question of whether the associated globular cluster population rotates \citep{2020MNRAS.491L...1L}.

There are several hypotheses that have been suggested to account for the distinct characteristics of these dark matter deficient galaxies.  \citet{2018MNRAS.473.4339O} proposes that these galaxies originally contained a typical amount of dark matter, which they later lost as a result of interactions with nearby galaxies, a scenario backed up through computer simulations
\citet{2022NatAs...6..496M}.  Based on this mechanism, they estimate that about 30\% of massive central galaxies contain a single satellite with little dark matter. The findings of \citet{2018anms.conf....1B} also suggest that the creation of some UDGs can be associated with interactions between galaxies, either due to UDGs coalescing from tidal debris in stellar streams.
Moreover, it is suggested by \citet{2022Natur.605..435V} that NGC 1052-DF2 and NGC 1052-DF4, the two dark matter deficient galaxies which are the focus of this study, originated in an individual event, a ``bullet dwarf'' collision, roughly eight billion years ago.

Recently, multiple investigations have confirmed that \df4 is experiencing tidal stripping, which may account for the galaxy's low dark matter content. \citet{2020ApJ...904..114M} showed that NGC 1052-DF4 is experiencing tidal disruption, and they infer that the contact between it and the nearby galaxy, NGC 1035, is most likely the cause for the stripping of dark matter. \citet{2022ApJ...935..160K} also discovered evidence of tidal disruption observed in both NGC 1052-DF2 and NGC 1052-DF4, which strongly suggests a common origin for the removal of dark matter.
However, they concluded that the most probable cause of these tidal disruptions is the central giant elliptical galaxy, NGC 1052, which lies between the two galaxies. 
More recently, \citet{2024arXiv240204304G} used ultra-deep images from the Gemini telescopes to explore tidal signatures in NGC 1052-DF2 and NGC 1052-DF4, finding no signs of tidal disruption, although NGC 0152-DF4 does appear to exhibit tidal tails. They, too, conclude that gravitational interactions may have removed the dark matter from NGC1052-DF4.

However, it is important to note that the claim of dark matter deficiency is
based on dynamical modeling that assumes no rotation \citep{2020MNRAS.491L...1L}. Such rotation shifts the balance between pressure and rotational support, and hence, neglecting it will bias the inferred dark matter mass
 \citep{2018ApJ...863L..15W,2022EPJC...82..935L}.Inspired by the work of \citet{2020MNRAS.491L...1L} on NGC 1052-DF2, a similar statistical analysis of NGC 1052-DF4's globular clusters was undertaken. The primary objective of this study is to determine whether the globular cluster population rotates.

The paper is structured as follows. Section~\ref{Data} describes the data for the NGC 1052-DF4 globular cluster population. Section~\ref{MandB} presents the kinematic model employed in this paper, and Bayesian inference is also explained in this section. Bayesian inference was used to estimate the posterior probability distributions and the marginal likelihoods for various kinematic models of the globular clusters. Section~\ref{result} presents the findings, while Section~\ref{Mass} presents the effect on the estimated mass of \df4. Finally, the discussion and conclusions are presented in Sections~\ref{Discussion} and~\ref{Conclusion}, respectively.

\section{Data}
\label{Data}



\begin{figure}[hbt!]
	\centering 
	\includegraphics[width=1\linewidth]{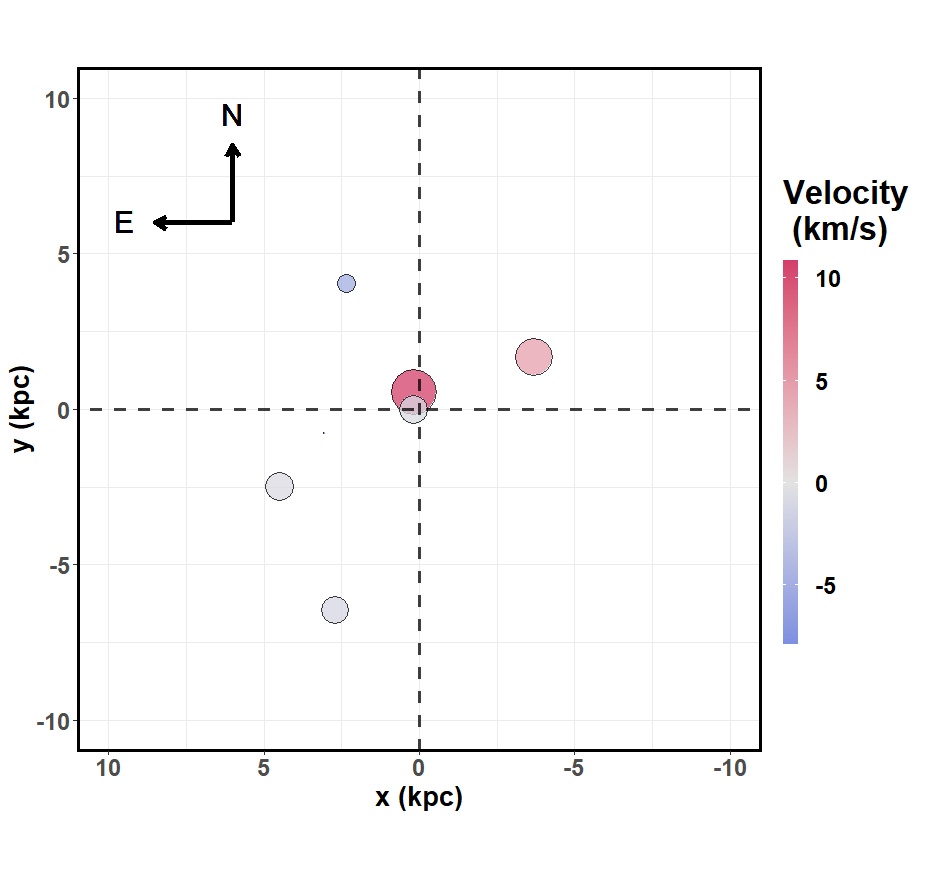}	
	\caption{Positions of the NGC 1051-DF4 globular clusters relative to the center of the galaxy. The size and colour of the circle represent each globular cluster's velocity along the line of sight; the larger the circle, the higher the absolute value of the velocity.} 
	\label{df4}%
\end{figure}

Previous authors used the Hubble Space Telescope
to discover the compact objects associated with NGC 1052-DF4 (van Dokkum et al. 2019). Using the technique described in \citet{2018ApJ...856L..30V}, SExtractor \citep{1996AAS..117..393B} was used to assess the overall magnitudes, colors, and Full Width Half Maximum (FWHM) sizes of compact objects in the images. \citet{2018ApJ...856L..30V} stated that compact objects with an $I_{814}$ magnitude smaller than 23, a $V_{606}-I_{814}$ colour  within the range between 0.20 and 0.43, and a FWHM value between 0.12'' and 0.30'' are considered to be likely globular clusters. Hence, by applying this method, \citet{2019ApJ...874L...5V} found that, out of 11 candidates, only seven meet these criteria, with a mean overall magnitude with a $I_{814}$ value of 22.10 and an rms spread of 0.39 mag. These seven objects were subsequently
confirmed as being associated with \df4 using spectra from the Low Resolution Imaging Spectrograph on the Keck I telescope.
Information about the seven confirmed globular clusters of NGC 1052-DF4 was available from \citet{2019ApJ...874L...5V}. 
For our purposes, the data consists of the positions of the clusters on the sky, the measured
velocities of the clusters along the line of sight, and the uncertainties on these line-of-sight velocities.
The positions and velocities of these seven clusters are shown in Figure~\ref{df4}.

\section{Models and Bayesian Inference}
\label{MandB}
 
The kinematic model used to analyze the NGC1052-DF2 galaxy \citep{2020MNRAS.491L...1L} is still used in this study. The line-of-sight velocity due to rotation is given by:
\begin{equation}\label{eqModel}
v_{r}(\theta) = A \sin(\theta - \phi),
\end{equation} where $A$ signifies the rotational amplitude and $\phi$ denotes how the rotation axis of the globular clusters is oriented.
This choice of functional form for the rotational velocity has previously been called
the $V$ model, after \citet{2014MNRAS.442.2929V}.
In other work, consideration was also given to the other two types of models, called $F$ and $S$. However, in order to maintain continuity with the previous analysis of NGC 1052-DF2, we chose to employ model $V$ in this analysis. \ref{Models} presents the details of the $F$ and $S$ models.

Bayesian inference is used throughout this work to determine the parameters' posterior probability distributions based on the data analysed. The posterior distribution is provided by Bayes's theorem:
\begin{equation}\label{br}
\mathcal{P}(\omega \,|\, \mathcal{D}) = \frac{\mathcal{P}(\omega)\mathcal{P}(\mathcal{D}\,|\,\omega)}{\mathcal{P}(\mathcal{D})},
\end{equation} where $\omega$ is a vector of unknown parameters for which an inference is anticipated (since $\phi$ and $\theta$ are used in this work to signify  angles on the sky, $\omega$ was chosen to denote unknown parameters). The data is represented by $\mathcal{D}$, the parameters' prior probability distribution is defined by $\mathcal{P}(\omega)$, $\mathcal{P}(\mathcal{D}|\omega)$ is the likelihood function, and $\mathcal{P}(\mathcal{D})$ is the marginal likelihood value, sometimes called the evidence. The prior distributions for unknown parameters are shown in Table \ref{t2.1}.

\begin{table}[hbt!]
\begin{threeparttable}
\caption{Table of prior probability distributions for the unknown parameters.}
\label{t2.1}
\begin{tabular}{llll}
\toprule
\headrow Parameters & Description  & Prior & Unit\\
\midrule
$A$ & Rotational Amplitude & Uniform(0, 20) & \kms   \\[0.5em]
\midrule
$\phi$ & Orientation of rotation axis & Uniform(0, 2$\pi$) & radians \\[0.5em]
\midrule
 $\sigma$ & Velocity dispersion & Uniform(0, 20)  & \kms\\[0.5em]
\midrule
$\vsys$ & Systemic Velocity & Uniform(-10, 10) & \kms\\[0.5em]
\bottomrule
\end{tabular}
\end{threeparttable}
\end{table}

The measured velocities $v_{i}$ given the parameters are assumed to have a Normal Distribution with mean $v_{r}(\theta)$ and dispersion of $\sigma^2 + \sigma_{i}^2$ as shown below:

\begin{equation}\label{eqLik}
v_{i}|\omega \sim \textnormal{Normal}(v_{r}(\theta) + \vsys, \sigma^2 + \sigma_i^2),
\end{equation}
so the likelihood function can be expressed as:
\begin{equation}\label{eqlik}
\mathcal{L}(\omega) = \prod_i \frac{1}{\sqrt{2\pi(\sigma^2+\sigma_i^2)}} \exp \left(-\frac{(v_i -(v_r(\theta)+\vsys)^2}{2(\sigma^2 + \sigma_i^2)}\right).
\end{equation}

The marginal likelihood or evidence is estimated using nested sampling introduced by \citet{2004AIPC..735..395S} and expressed as follows:
\begin{equation}\label{NS}
\mathcal{Z}  = \int \mathcal{L}(\omega) \times \pi (\omega) \,d\omega,
\end{equation} where $\mathcal{Z}$ is the evidence (marginal likelihood), $\mathcal{L}(\omega)$ represents the likelihood function, and $\pi(\omega)$ stands for the prior distribution. Since nested sampling computes the marginal likelihood for each model constructed, we are able to choose the most plausible model based on the highest marginal likelihood value (assuming equal prior probabilities) or propagate uncertainty about the model into any conclusions reached.

The Bayes factor is subsequently computed to assess which model fits the data best after obtaining the marginal likelihood for each model. It is a ratio of the marginal likelihood of two models, and its value determines how strongly the data supports one model over another. The Bayes Factor is given by:
\begin{equation}\label{BF}
\mathcal{B}\mathcal{F}(\mathcal{M}_1,\mathcal{M}_2) = \frac{\mathcal{P}(\mathcal{D}\,|\,\mathcal{M}_1)}{\mathcal{P}(\mathcal{D}\,|\,\mathcal{M}_2)},
\end{equation}
where the model is denoted by $\mathcal{M}_i$.

\section{Results}
\label{result}

\subsection{Model 1: Non-Rotational Model}
\label{M1}
\begin{figure}[hbt!]
	\centering 
	\includegraphics[width=1\linewidth]{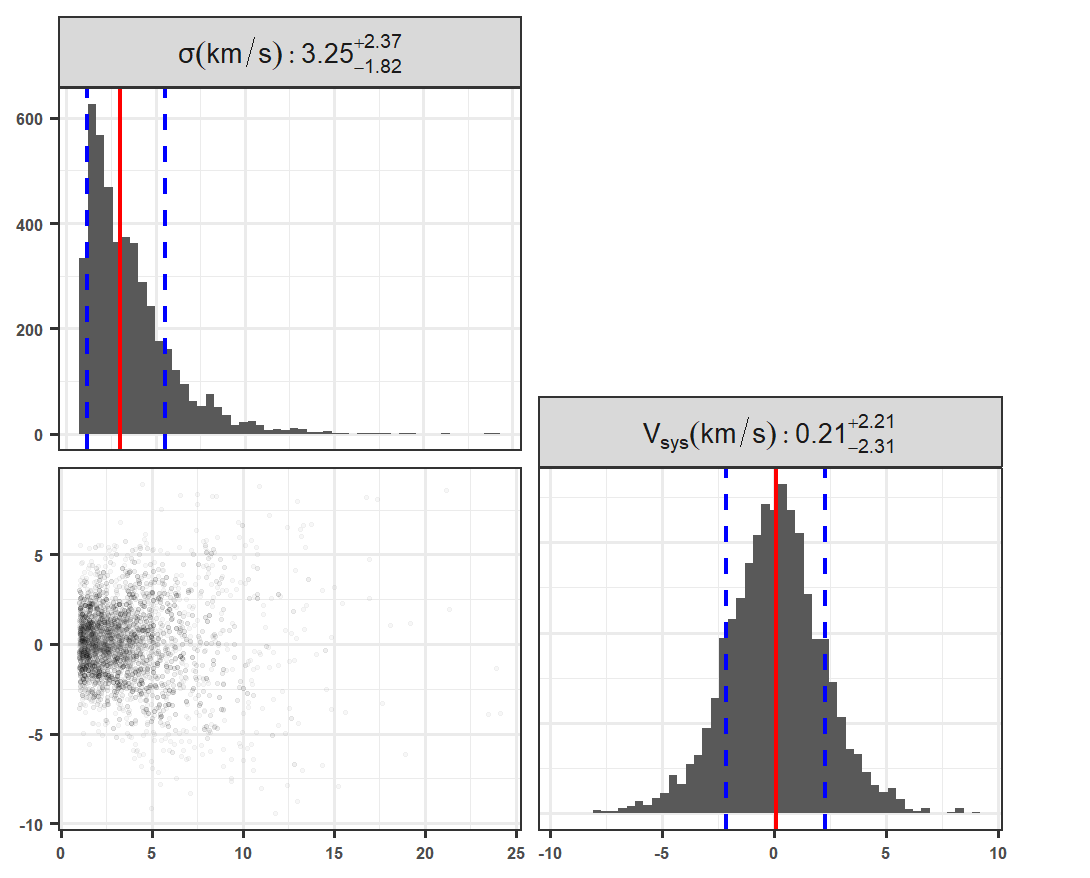}	
	\caption{Corner plot of the parameters' posterior distribution for the Non-Rotational Model.} 
	\label{nr}%
\end{figure}

We first consider a scenario in which the globular cluster population does not rotate.
Under this assumption, the amplitude $A$ in Equation \ref{eqModel} will be equal to zero, and hence the velocity $v_{i}$ is entirely dependent on the velocity dispersion $\sigma$. This leaves only two parameters to be estimated, $\sigma$ and $v_{\rm sys}$. Table \ref{t2.2} displays the summary statistics of parameter estimates, and Figure \ref{nr} shows the posterior distributions for systematic velocity $\vsys$ and velocity dispersion $\sigma$. The distribution of $\sigma$ is right-skewed, whereas the distribution of $\vsys$ is symmetric.
Finally, this Non-Rotational Model has a marginal likelihood of $\ln{Z} = -25.6095 \pm 0.1762$.

\begin{table}[hbt!]
\begin{threeparttable}
\caption{Posterior summary statistics of parameters for the Non-Rotation Model. The estimates presented in the table were the median value and the 68\% central credible interval. All of the values were rounded to 2 d.p.}
\label{t2.2}
\begin{tabular}{lll}
\toprule
\headrow Parameters & Estimates  & Units\\
\midrule
$\sigma$ & $3.25\quad ^{+2.37}_{-1.82}$ & \kms \\[0.5em]
\midrule
$\vsys$ & $0.21\quad ^{+2.21}_{-2.31}$ & \kms\\[0.5em]
\bottomrule
\end{tabular}
\end{threeparttable}
\end{table}

\subsection{Model 2: Rotational Model}
\begin{figure*}
\centering
\includegraphics[width=0.65\linewidth]{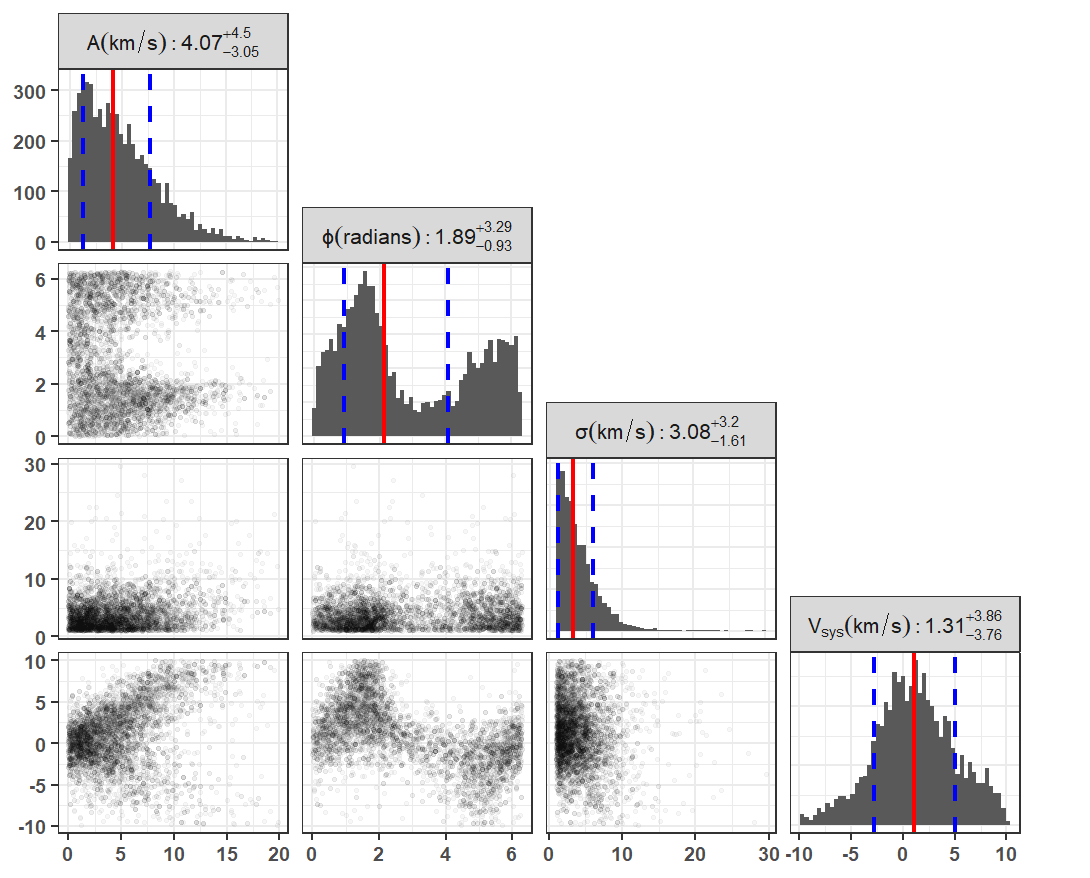}
\caption{Corner plot of the parameters' posterior distribution for the Rotational Model.} 
\label{r}
\end{figure*}

The second model we considered includes a nonzero rotational amplitude $A$.
The posterior distribution for the parameters of the Rotational Model is displayed in Figure~\ref{r}. There is a right-skewed posterior distribution of parameters $A$ and $\sigma$, with most probabilities accumulating between 0 \kms and 7 \kms and 0 \kms and 6 \kms, respectively. Furthermore, the posterior distribution of $\vsys$ is symmetric. Interestingly, the posterior distribution for $\phi$ is weakly bimodal, as there are multiple ways of dividing the globular clusters into a red-shifted half and a blue-shifted half (on average). Note that the apparent large bimodality is an artifact of the periodicity of the parameter space --- the smaller, actual bimodality can be seen when angles are redefined to be centered around zero.
The mean and median values of amplitude, $A$, are 4.88 \kms and 4.07 \kms, which are somewhat higher than those of velocity dispersion, $\sigma$. The summary statistics of parameter estimates are shown in Table \ref{t2.3} and the marginal likelihood of the Rotational Model is $\ln{Z} = -26.5079 \pm 0.1919$.

\begin{table}[hbt!]
\begin{threeparttable}
\caption{Posterior summary statistics of parameters for the Rotational Model. The estimates presented in the table were the median value and the 68\% central credible interval. All of the values were rounded to 2 d.p.}
\label{t2.3}
\begin{tabular}{lll}
\toprule
\headrow Parameters & Estimates  & Units\\
\midrule
$A$ & $4.07\quad ^{+4.50}_{-3.05}$ & \kms\\[0.5em]
\midrule
$\phi$ & $1.89\quad ^{+3.29}_{-0.93}$ & radians\\[0.5em]
\midrule
$\sigma$ & $3.08\quad ^{+3.20}_{-1.61}$ & \kms\\[0.5em]
\midrule
$\vsys$ & $1.31\quad ^{+3.86}_{-3.76}$ & \kms\\[0.5em]
\bottomrule
\end{tabular}
\end{threeparttable}
\end{table}

\subsection{Model 3: Rotational Model with Alternative Priors}
Finally, we considered the rotational model but with a more complex joint prior distribution for amplitude $A$, velocity dispersion $\sigma$, and systemic velocity $\vsys$,
rather than the simple uniform distributions used above. These priors model the idea that the rotational amplitude is likely not to be precisely zero but may be small, moderate, or large. Roughly speaking, the rotating and non-rotating (or, more accurately, hardly-rotating) possibilities are here represented in the parameter space of a single parameter estimation problem, rather than being represented in two separate models, as we have done so far. The prior distribution stays the same for the orientation of the rotation axis $\phi$. Table \ref{t2.4} presents the details of the alternative prior distribution of the parameters.

\begin{table*}[h]
\begin{threeparttable}
\caption{Prior distributions for the parameters in the Rotational Model with Alternative Priors.}
\label{t2.4}
\begin{tabular}{llll}
\toprule
\headrow Parameters & Description  & Prior & Unit\\
\midrule
$A$ & Rotational Amplitude & $(10^{1-|\textnormal{Student}-t(0, 0, 2)|})\sigma$ & \kms \\[0.5em]
\midrule
$\phi$ & Orientation of rotation axis &  Uniform(0, 2$\pi$) & radians \\[0.5em]
\midrule
$\log_{10}\sigma$ & Velocity dispersion & $\textnormal{Student}-t(1, 0.5, 4)$ & \kms\\[0.5em]
\midrule
$\vsys|\sigma$ & Systemic Velocity & $\textnormal{Student}-t(0, 0.1\sigma, 1)$ & \kms\\[0.5em]
\bottomrule
\end{tabular}
\end{threeparttable}
\end{table*}

\begin{figure*}
\centering
\includegraphics[width=0.65\linewidth]{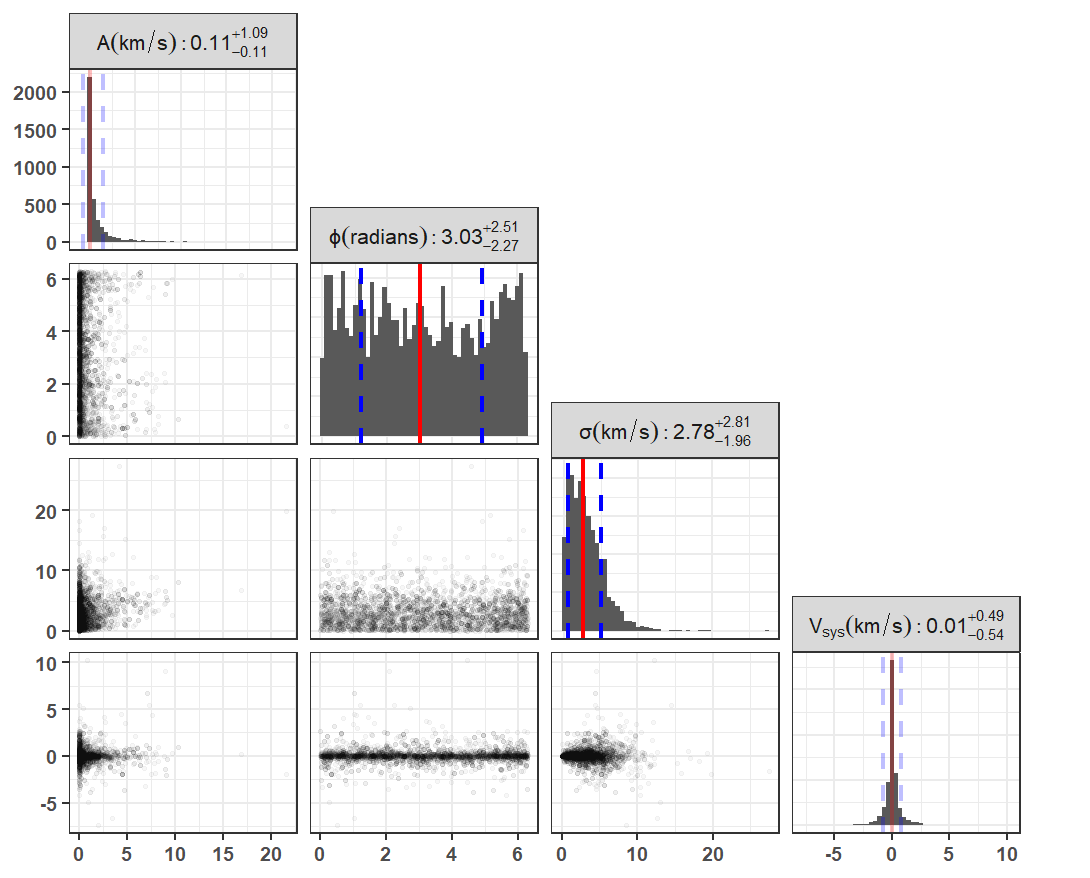}
\caption{Corner plot of the parameters' posterior distribution for the Rotational Model with Alternative Priors.} 
\label{ar}%
\end{figure*}

Figure \ref{ar} demonstrates a similar outcome to Figure \ref{r}; the posterior distribution of parameters $A$ and $\sigma$ is right-skewed, with the majority of the probabilities gathering between 0 kms$^{-1}$ and 2 kms$^{-1}$ and 0 kms$^{-1}$ and 6 kms$^{-1}$, respectively. Additionally, the posterior distribution of $\vsys$ shows that it is effectively zero (though it will never be precisely zero) and is still symmetric. Compared to the outcome of the Rotational Model, the amplitude, $A$, from this model, has a much lower value for the mean and median and is smaller than those values for the velocity dispersion, $\sigma$. In this case, the posterior distribution for $\phi$ is approximately equal to the prior distribution, as the angle becomes less important when the amplitude $A$ is low, as it is here. The marginal likelihood of the Rotational Model with Alternative Priors is $\ln{Z} = -23.0067 \pm 0.1014$. Table \ref{t2.5} presents parameter estimation summary statistics.

\begin{table}[hbt!]
\begin{threeparttable}
\caption{Posterior summary statistics of parameters for the model with alternative priors. The estimates presented in the table were the median value and the 68\% central credible interval. All of the values were rounded to 2 d.p.}
\label{t2.5}
\begin{tabular}{lll}
\toprule
\headrow Parameters & Estimates   & Unit\\
\midrule
$A$ & $0.11\quad ^{+1.09}_{-0.11}$ & \kms \\[0.5em]
\midrule
$\phi$ & $3.03\quad ^{+2.51}_{-2.27}$ & radians\\[0.5em]
\midrule
$\sigma$ & $2.78\quad ^{+2.81}_{-1.96}$ & \kms \\[0.5em]
\midrule
$\vsys$ & $0.01\quad ^{+0.49}_{-0.54}$ & \kms\\[0.5em]
\bottomrule
\end{tabular}
\end{threeparttable}
\end{table}

\subsection{Model Comparison}
\label{MC}
The study will not compare the marginal likelihood of Model 3 (Rotational Model with  Alternative Priors) to the other two models since it is intended to capture both possibilities within a single model. 
Table \ref{t2.6} shows that the marginal likelihood estimate for the Non-Rotational model is slightly larger than the Rotational Model. The Bayes Factor of the Non-Rotational Model over the Rotational Model is calculated as below:
\begin{equation}\label{eqBF}
    \begin{split}
     \mathcal{B}\mathcal{F}(\mathcal{M}_1, \mathcal{M}_2) & = \frac{\exp(\log\mathcal{P}(\mathcal{D}\,|\,\mathcal{M}_1))}{\exp(\log\mathcal{P}(\mathcal{D}\,|\,\mathcal{M}_2))}\\ 
     & = \frac{\exp(-25.6095)}{\exp(-26.5079)} \approx 2.46,
    \end{split}
\end{equation}
where, $\mathcal{M}_1$ is the Non-Rotational Model and $\mathcal{M}_2$ is the Rotational Model.
This Bayes Factor of around 2.5 indicates weak support for the non-rotational model.

\begin{table}[hbt!]
\begin{threeparttable}
\caption{Table of Marginal likelihood estimates for the Non-Rotational and Rotational Model. All of the values were round to 2 d.p.}
\label{t2.6}
\begin{tabular}{ll}
\toprule
\headrow Models & $\ln{Z}$\\
\midrule
Non-Rotational & $-25.61\ \pm \ 0.18$\\[0.5em]
\midrule
Rotational & $-26.51\ \pm \ 0.19$\\[0.5em]
\bottomrule
\end{tabular}
\end{threeparttable}
\end{table}
\section{Mass of the Galaxy}
\label{Mass}

As stated in the introduction, the idea that the galaxy is dark matter deficient is supported by a mass estimate based on the assumption of no rotation. If a rotational signature is detected in the globular clusters in Section \ref{result}, the estimated dynamical mass of NGC 1052-DF4 will be affected, which would further affect conclusions about the quantity of dark matter in the galaxy. Therefore, the study investigated employing the same estimator that \citet{2020MNRAS.491L...1L} used to estimate the overall mass of the NGC 1052-DF2 galaxy within a certain radius. The estimator is expressed as follows:
\begin{equation}\label{eqmass}
M( < r) = \left( \left( \left(\frac{v_{rot}}{\sin(i)} \right)^2 + \sigma^2\right)\right) \frac{r}{G},
\end{equation}
where, $v_{rot}$ is the rotational velocity as given in equation \ref{eqModel}, $\sigma$ for velocity dispersion, 
and according to the standard astronomical definition, $i$ is the rotation's 
inclination angle. In this section, the amplitude, $A$, serves as the only representation of
the rotating velocity, $v_{\rm rot}$. Moreover, $r$ is the reference radius with a value of
7.5 kpc, and $G$ is the Newtonian gravitational constant.
According to \citet{2009ApJ...704.1274W} and \citet{2010MNRAS.406.1220W}, equation \ref{eqmass} can be considered as a lower estimate
of the galaxy's total mass because the more complicated mass estimators normally have a
multiplication constant together with the square of the $\sigma$, which explains the
internal mass distribution, and usually exceeds one. The mass estimator here is a function of the unknown parameters. To compute the posterior distribution for the mass, we applied the formula to each possible parameter vector in our posterior sample.

\begin{figure*}
	\centering 
	\includegraphics[width=0.65\linewidth]{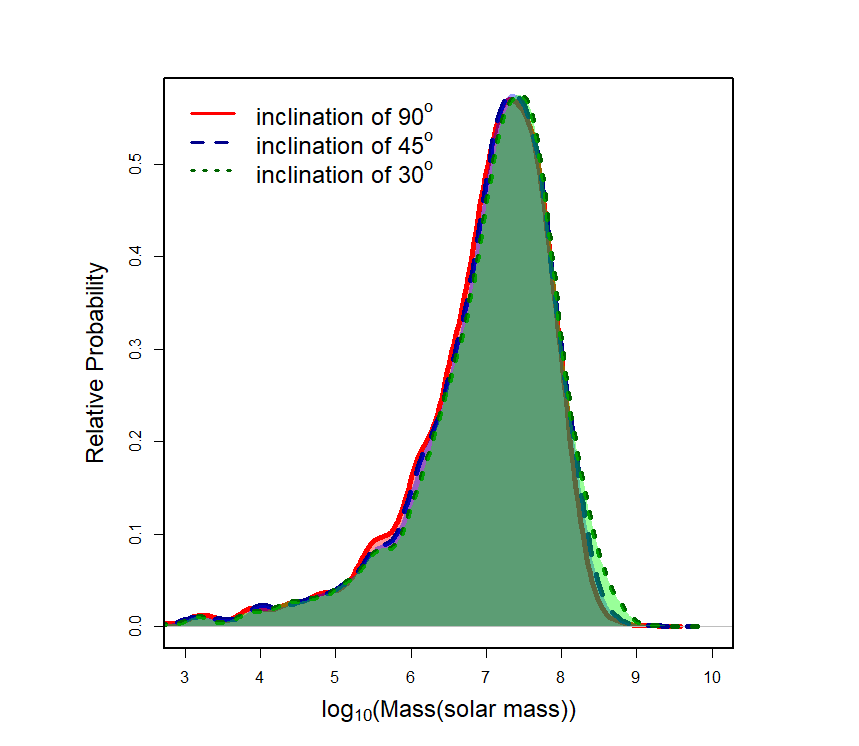}	
	\caption{Posterior distribution of the estimated mass of NGC 1052-DF4 with the different rotational inclination. The red curve represents a $90^o$  inclination, the blue dash curve has a $45^o$  inclination, and the green dot curve has a $30^o$  inclination.} 
	\label{density}%
\end{figure*}

The posterior distribution of NGC 1052-DF4's estimated mass with three different rotational inclinations is shown in Figure \ref{density}. This figure suggests that the estimated mass of NGC 1052-DF4 has right-skewed posterior distributions for all three inclination angles, with the majority of the probability falling between 0 and $10^8$ solar masses. Moreover, the estimated mass of the NGC 1052-DF4 can reach as high as 6$\times 10^8$ solar masses. There is no substantial difference between the three density curves with different inclinations. This is reasonable as the amplitude, $A$, was very small or possibly zero, which caused the velocity dispersion, $\sigma$, to dominate in the equation \ref{eqmass} and result in the inclination becoming irrelevant.

\begin{figure*}
	\centering 
	\includegraphics[width=0.65\linewidth]{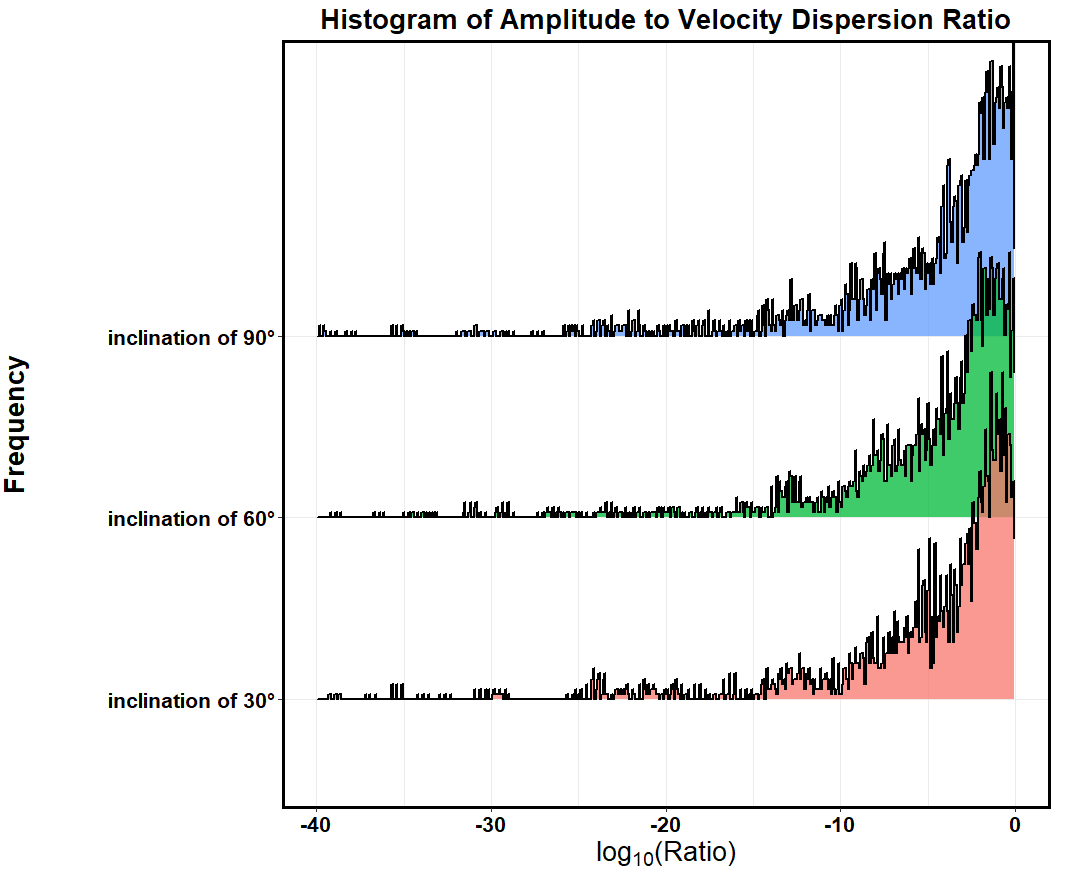}	
	\caption{Histograms of Amplitude to Velocity Dispersion Ratio with different rotational inclination. The first histogram figure displays a ratio distribution with a $90^o$ inclination, the second histogram plot illustrates a ratio distribution with a $45^o$ inclination, and the final histogram plot shows a distribution with a $30^o$  inclination. 
} 
	\label{ratio}%
\end{figure*}

Figure \ref{ratio} displays the posterior distributions of the log amplitude-to-velocity-dispersion ratio, $A/\sigma$, with different inclinations. The distributions of $A/\sigma$ with different inclinations are all remarkably similar, with the majority of the probability being for large negative values of the log ratio. This demonstrates that the NGC 1052-DF4 galaxy's globular cluster population has a low probability of containing a significant rotating component, and that estimates of its mass do not need to take any significant rotation into account.

\section{Discussion}
\label{Discussion}

Through the fitting of three Bayesian models to the data, the rotational characteristics of the globular cluster population of NGC 1052-DF4 have been assessed. Standard uniform priors were used for the Non-Rotational Model and one of the Rotational Models, and under these assumptions, we performed a model comparison, finding that the Non-Rotational Model was favoured. As an alternative, using a more complex prior, we performed a similar analysis using only parameter estimation. In this latter case, the question being answered changed from ``Is the amplitude zero or non-zero?" to ``Is the amplitude small or large?"

The posterior distribution in Figure \ref{r} reveals that a significant portion of the posterior probabilities for the rotational amplitude, $A$, and velocity dispersion, $\sigma$, are close to zero. This implies the rotational amplitudes of the globular clusters in NGC 1052-DF4 are extremely small. Moreover, the right-skewed posterior distribution of $\sigma$ results from the small sample size, as high values of $\sigma$ cannot necessarily be ruled out. Model 3 (Rotational Model with Alternative Priors) had a similar result for the amplitude of rotation, $A$, and velocity dispersion, $\sigma$. The 68\% central credible interval for $A$ extends from 0.00 kms$^{-1}$ to only 1.19 kms$^{-1}$. As can also be seen from Table \ref{t2.3}, both of the estimations of the rotational amplitudes, $A$, are smaller than those for the velocity dispersion, $\sigma$. 

As seen in Table \ref{t2.6}, the Non-Rotational Model's marginal likelihood is greater than that of the Rotational Model. Although the difference between the two is small, it does indicate that the data favor the Non-Rotational Model over the Rotational Model. However, according to the value (2.46) of the Bayes factor calculated in Equation \ref{eqBF}, there is only weak evidence to support the Non-Rotation Model (guides for interpretation of Bayes Factors may be found in \citet{kass} and \citet{PENNY2007454}). In other words, the rotation of NGC 1052-DF4's globular clusters is not completely ruled out. However, if it does exist, it is small. 

Based on equation \ref{eqModel}, the amplitude is directly proportional to the rotational velocity. Also, from equation \ref{Mass}, it can be seen that the rotational velocity and the velocity dispersion, $\sigma$, are directly related to the mass of the galaxy. Consequently, a low amplitude value (around 0) and a slightly larger value of $\sigma$ indicate that the globular cluster rotational velocities will likewise be small, resulting in a very low estimated mass of the NGC 1052-DF4 galaxy, which is about the same as in the previous study \citep{2019ApJ...874L...5V}. Hence, these results further imply that the NGC 1052-DF4 galaxy contains little dark matter, which is consistent with the result of a newly published paper from \citet{2023ApJ...957....6S}. As mentioned in Section \ref{introduction}, \citet{2024arXiv240204304G} found that NGC 0152-DF4 exhibits tidal tails, which explains the low quantity of dark matter in the galaxy. 


\section{Conclusions}
\label{Conclusion}

In this paper, we tested the existence of a rotating element by analysing the rotational features of the seven globular clusters in NGC 1052-DF4. After comparing the marginal likelihood and calculating the Bayes factor of the Non-Rotational Model versus the Rotational Model, we find that the Non-Rotational Model is weakly preferred by a Bayes Factor of about 2.5. The outcome of the Rotational Model with Alternative Priors provides more evidence that the globular cluster populations of \df4 do not rotate significantly. The result from this study is consistent with the previous study \citep{2019ApJ...874L...5V}. Furthermore, this result supports the conclusion that there is little dark matter in the NGC 1052-DF4 galaxy since mass estimates to not need to be modified to take rotation into account. This result is also consistent with the recent study from \citet{2024arXiv240204304G} and reinforces the conclusion from \citet{2023ApJ...957....6S}. 


\begin{acknowledgement}
We thank the anonymous referee for helpful comments which assisted us in improving this manuscript. 
\end{acknowledgement}

\paragraph{Funding Statement}
This research received no external funding.
\paragraph{Competing Interests}

None

\paragraph{Data Availability Statement}

Any reasonable request will be granted access to the data used in this work.

\printendnotes
\printbibliography

\appendix

\section{Other Kinematic Models}
\label{Models}

In this paper, we considered applying three different types of models (Table \ref{FSV}) to analyze the rotational features of NGC 1052-DF4's globular cluster population:

\begin{itemize}
    \item 
    \citet{2014MNRAS.442.2929V} introduced the model $V$. This model takes into account a rotational velocity with a constant amplitude that is affected by angular dependence.
    \item
    In the $S$ model, the globular cluster population rotates like a solid body, and the velocity increases linearly with respect to the rotation axis.
    \item
    According to model $F$, the rotation of the globular clustersis asymptotically flat. It is predicted that the velocity approaches a constant value away from the rotation axis, comparable to the rotation curve of spiral galaxies.
\end{itemize}

\begin{table}[H]
\begin{threeparttable}
\caption{Table of the three models' expressions for rotational velocity.}
\label{FSV}
\begin{tabular}{ll}
\toprule
\headrow Models & functional form of the models\\
\midrule
$V$  & $v_{r}(x_i, y_i) = A \sin(\theta - \phi)$ \\[0.5em]
\midrule
$S$  & $v_{r}(x_i, y_i) = A (x\sin(\phi) - y\cos(\phi))$ \\[0.5em]
\midrule
$F$  & $v_{r}(x_i, y_i) = A \tanh((x\sin(\phi) - y\cos(\phi))/L)$ \\[0.5em]
\bottomrule
\end{tabular}
\end{threeparttable}
\end{table}

Although Model $F$ had the highest marginal likelihood estimates for both the Rotation Model and the Rotation Model with Alternative Priors, the difference is incredibly small when compared to the marginal likelihood estimates from Model $V$ for both the Rotation Model and the Rotation Model with Alternative Priors (Table \ref{fsv}). However, for consistency with the earlier study of NGC 1052-DF2, which used Model $V$, we have decided to use model $V$ for this study. The main conclusions of the paper are not sensitive to the choice of rotation model.

\begin{table}[H]
\begin{threeparttable}
\caption{Table of Marginal Likelihood Estimates for the Models. Here, model 2 is represented as the Rotation Model, and Model 3 is the Rotation Model with Alternative Priors. Model $F$ is the one that the data favors the most out of the three models. Because the study assumes the globular clusters are not rotating in the Non-Rotation Model, the comparison of the Non-Rotation Model among the three models is meaningless. As a result, amplitude A will be zero in all $V$, $S$, and $F$ models, providing the same outcomes for all three models. All of the values were rounded to 2 d.p.}
\label{fsv}
\begin{tabular}{lll}
\toprule
\headrow Models & $\ln{Z}$ for Model 2 & $\ln{Z}$ for Model 3\\
\midrule
$V$  & $-26.51 \pm 0.19$ & $-23.00 \pm 0.10$\\[0.5em]
\midrule
$S$  & $-27.53 \pm 0.24$ & $-22.99 \pm 0.11$\\[0.5em]
\midrule
$F$  & $-26.25 \pm 0.21$ & $-22.92 \pm 0.11$\\[0.5em]
\bottomrule
\end{tabular}
\end{threeparttable}
\end{table}

\end{document}